\documentclass{article}
\usepackage{amsmath}
\usepackage{amssymb}
\usepackage{amsthm}
\usepackage{latexsym}
\usepackage{epsfig}
\usepackage{url}
\DeclareMathOperator{\conv}{conv}
\begin{document}
\newcommand{\ontop}[2]{\genfrac{}{}{0pt}{}{#1}{#2}}

\title{Parametric Alignment of Drosophila Genomes}

\author{Colin Dewey, Peter Huggins, \\
Kevin Woods, Bernd Sturmfels, Lior Pachter \\ \\ Department of
Mathematics
\\ University of California at Berkeley}

\date{ }
\maketitle

 \begin{abstract}

 The classic algorithms of Needleman--Wunsch and Smith--Waterman find
 a {\em maximum a posteriori} probability alignment for a pair hidden Markov
 model (PHMM). In order to process large genomes that have undergone 
complex genome rearrangements, almost all
  existing whole genome alignment methods
 apply fast heuristics to divide genomes into small pieces
 which are suitable for Needleman--Wunsch alignment.
 In these alignment methods, it is
standard practice to fix the
 parameters and to produce a single alignment for subsequent
 analysis by biologists.

 As the number of alignment programs applied on a whole genome scale
 continues
 to increase, so does  the disagreement in their results.
 The alignments produced by different programs vary greatly, especially
 in non-coding regions of eukaryotic genomes where
 the biologically correct  alignment is hard to find.
 Parametric alignment is one possible remedy.
 This methodology resolves the issue of robustness to  changes in parameters
 by finding all optimal alignments for all possible parameters in a PHMM.

 Our main result is the construction of a whole genome parametric alignment
 of \textit{Drosophila melanogaster} and \textit{Drosophila pseudoobscura}.
 This alignment draws on existing heuristics for dividing whole genomes
 into small pieces for alignment, and it relies on advances
 we have made in computing convex polytopes that allow us to
 parametrically align non-coding regions using biologically
 realistic models. We demonstrate the utility of our parametric alignment
 for biological inference by  showing that cis-regulatory
elements are more conserved between {\it Drosophila melanogaster} and
{\it Drosophila pseudoobscura} than previously thought. We also 
show how whole genome parametric alignment can be used to
quantitatively assess the dependence of branch length estimates on
alignment parameters. 

The alignment polytopes, software, and supplementary material can be
downloaded at \ {\tt http://bio.math.berkeley.edu/parametric/}.
 \end{abstract}

 \newpage
 \section{Introduction}
\label{Sect:Introduction}

 Needleman--Wunsch pairwise sequence alignment \cite{Needleman1970}
 is known to be sensitive to parameter choices. To illustrate the problem,
 consider the 8th intron of the \textit{Drosophila melanogaster} CG9935-RA
 gene (as annotated by FlyBase \cite{Drysdale2005}) located on chr4:660,462-660,522 (April 2004 BDGP release 4).
 This intron, which is 61 base pairs long, has a 60 base pair ortholog
in {\it Drosophila pseudoobscura}. The ortholog is located
  at Contig8094\_Contig5509:4,876-4,935 in the
August 2003, freeze 1 assembly, as produced by the
Baylor Genome Sequencing Center.

 Using the basic 3-parameter scoring scheme (match $M$, mismatch $X$
 and space penalty $S$),   these two orthologous introns
 have the following optimal alignment  when the parameters are set to
 $M=5$, $X=-5$ and $S=-5$:
 \begin{small}
 \begin{verbatim}
mel GTAAGTTTGTTTAT-ATTTTTTTTTTTTTGAAGTGA-CAAATAGC-A-CTTATAAATATACTTAG
pse GTTCGTTAACACATGAAATTCCATCGCCTGAT-TGTTCA-CTATCTAACTAACGAAT-T--TTAG
    **  ***     ** *  **   *    ***  **  **  ** * * ** *  *** *  ****
       \end{verbatim}
 \end{small}
\vskip -.4cm
 However, if we change the parameters to $M=5$, $X=-6$ and $S=-4$,
 then the following  alignment is optimal:
 \begin{small}
 \begin{verbatim}
mel GTAAGTT------TGTTTATATTTTTTTT--T--TT-TTGAAGTGA-CAAATAGCACTTATA--A
pse GTTCGTTAACACATG-A-A-ATTCCATCGCCTGATTGTT-CACT-ATC---TA--AC-TA-ACGA
    **  ***      **   * ***   *    *  ** **  * * * *   **  ** ** *  *

mel ATATACTTAG
pse AT-T--TTAG
    ** *  ****
 \end{verbatim}
 \end{small}
 \vskip -.4cm
Note that a relatively small change in the parameters produces a
very different alignment of the introns.  This problem is exacerbated
 with more complex scoring schemes, and is a central
 issue with whole genome alignments produced by programs
 such as MAVID \cite{Bray2003} or BLASTZ/MULTIZ
 \cite{Schwartz2003}. Indeed, although whole
 genome alignment systems use many heuristics for rapidly identifying
 alignable regions  and subsequently aligning them,
 they all rely on the Needleman--Wunsch algorithm at some
 level. Dependence on parameters becomes an even more
 crucial issue in the multiple alignment of more than two sequences.

 {\em Parametric alignment} was introduced
 by Waterman, Eggert and Lander \cite{Waterman1992} and further
 developed by Gusfield et al. \cite{Gusfield1994, Gusfield1996}
 and Fernandez-Baca et al. \cite{Fernandez-Baca2004}
 as an approach for overcoming the difficulties
 in selecting parameters for Needleman--Wunsch alignment.
 See \cite{Fernandez-Baca2005} for a review and
 \cite{Pachter2004b, ASCB2005} for an algebraic perspective.
 Parametric alignment amounts to
 partitioning the space of parameters into regions.
 Parameters in the same region lead to the same
 optimal alignments. Enumerating all regions is
 a non-trivial problem of computational geometry.
 We solve this problem on a whole genome scale
 for up to five free paramaters.

 Our approach to parametric alignment rests on the
 idea that the score of an alignment is specified by
 a short list of numbers derived from the alignment. For instance, given the
 standard 3-parameter scoring scheme, we summarize
 each alignment by the number $m$ of matches,
 the number $x$ of mismatches,
 and the number $s$  of spaces in the alignment.
 The triple $(m,x,s)$  is called the {\em alignment summary}.
 As an example consider the above pair of orthologous {\it Drosophila} introns.
 The first (shorter) alignment has the   alignment summary $(33,23,9)$
 while the second (longer) alignment has the alignment summary
 $(36,10,29)$.

 \begin{table}
\label{sevenoptimal}
 $$
\begin{matrix} &
\hbox{alignment summary} &
\hbox{number of alignments with that summary} \\
{\bf A} & (25, 35, 1)  &  5 \\
{\bf B} & (28, 31, 3)   & 15 \\
{\bf C} & (32, 25, 7) & 44 \\
{\bf D} & (33, 23, 9)  & 78 \\
{\bf E} & (34, 20, 13)  &  156 \\
{\bf F} & (36, 10, 29) & 8064
\end{matrix}
$$
\vskip -.3cm \caption{The 8362 optimal alignments for two
{\it Drosophila} intron sequences.}
\end{table}

 Remarkably, even though the number of all alignments
 of two sequences is very large, the number of alignment
 summaries that arise from Needleman--Wunsch alignment is very small.
 Specifically, in the example above, where
 the two sequences  have lengths 61 and 60,
 the total number of alignments is
 \begin{small}
 $$
 \text{1,511,912,317,060,120,757,519,610,968,109,962,170,434,175,129}
 \, \simeq \,1.5 \times 10^{46}.
 $$
 \end{small}
There are  only 13 alignment summaries  that have the highest
score
 for some choice of parameters $M,X,S$. For
 biologically reasonable  choices, i.e., when we require
 $M > X$ and $2S < X$, only six of the 13 summaries are optimal.
 These six summaries account for a total of 8362 optimal alignments
(Table  \ref{sevenoptimal}).

Note that the basic model discussed above has
 only $d=2$ free parameters,  because
for a pair of sequences of lengths $l,l'$ all the summaries $(m,x,s$) satisfy
\begin{equation}
\label{linrel}
 2m+2x+s \,\,\, = \,\,\, \ell+\ell'.
 \end{equation}
 This relation holds with $\ell+\ell'=121$ for the six summaries 
  in Table \ref{sevenoptimal}.  Figure~1 shows the alignment polygon,
  as defined in Section 2.3, in the coordinates $(x,s)$.

\begin{figure}[ht]
\label{fig:polygon1}
\begin{center}
\includegraphics[scale=.40]{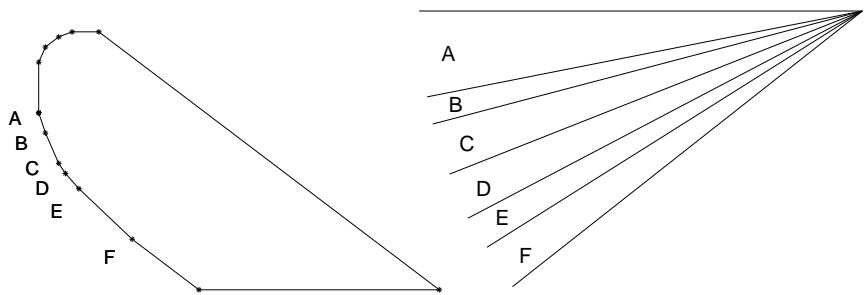}
\end{center}
\caption{The alignment polygon for our
two introns is shown on the left.
For each of the alignment summaries ${\bf A},{\bf B},\ldots,{\bf F}$ 
in Table 1, the corresponding cone in the alignment fan 
is shown on the right. If the
parameters $(S,X)$ stay inside a particular cone, every
optimal alignment has the same alignment summary.}
\end{figure}

 In general, for two DNA sequences of
 lengths $\ell$ and $\ell'$, the number of optimal alignment summaries
 is bounded from above by a polynomial in $\ell+\ell'$ of
 degree $d(d-1)/(d+1)$,
 where $d$ is the number of {\em free parameters} in the model
 \cite{Fernandez-Baca2005,  Pachter2004b}.
 For $d=2$, this degree is $0.667$, and so the number of optimal
alignment summaries has {\em sublinear growth} relative to the
sequence lengths. Even for $d=5$, the growth exponent
$d(d-1)/(d+1)$ is only $3.333$. This means that
 \emph{all} optimal alignment summaries can be
 computed on a large scale for models with few parameters.

The growth exponent $d(d-1)/(d+1)$ was derived by Gusfield et
al.~\cite{Gusfield1994} for $d=2$ and by Fernandez-Baca at
al.~\cite{Fernandez-Baca2004} and Pachter--Sturmfels
\cite{Pachter2004b} for general $d$. Table \ref{sevenoptimal} can be computed using
the software XPARAL \cite{Gusfield1996}. This software works for
$d=2$ and $d=3$, and it generates a representation of all
optimal alignments with respect to all reasonable choices of
parameters. Although XPARAL has a convenient graphical interface, it
seems that this program has not been widely used by biologists,
 perhaps because it is not designed for high throughput data
analysis and the number of free parameters is restricted to $d \leq
3$.

In this paper we demonstrate that parametric sequence alignment can be
made practical on the whole-genome scale, and we argue that
computing output such as Table \ref{sevenoptimal} can be very useful
for comparative genomics applications where reliable alignments
are essential. To this end, we introduce a mathematical point of
view, based on the organizing principle of {\em convexity}, that was
absent in the earlier studies \cite{Waterman1992, Gusfield1994,
Fernandez-Baca2005}. Our advances rely on new algorithms, which are
quite different from what is implemented in XPARAL, and which
perform well in practice, even if the number $d$ of free parameters
is greater than three.

 Convexity is the organizing principle that
 reveals the needles in the haystack.
 In our example, the ``haystack'' consists of more than $10^{46}$
 alignments, and the  ``needles'' are the 8362 optimal alignments.
 The summaries of the optimal alignments
are the vertices of the {\em alignment
 polytope}. The alignment polytope is the {\em convex hull} of the
 summaries  of all (exponentially many) alignments.
 Background on convex hulls and how to compute the
 alignment polytopes
 are provided in Section 2; see also
 \cite[\S 2.2]{ASCB2005}. Thus,
 parametric alignment of two DNA sequences
 relative to some chosen PHMM
 means  constructing the alignment polytope of the two sequences.
 The dimension of the alignment polytope  is $d$,
 the number of free model parameters.
For $d=2$ (the basic model),
 the polytope  is a convex polygon,
 as shown in  Figure~\ref{fig:polygon1}
 for the pair of introns above.

 The basic model is insufficient for
genomics applications. More realistic PHMMs
 for sequence alignment include 
gap penalties.  We consider three such
 models. The symmetries of the scoring matrices
for these models are derived from those of the evolutionary models
known as Jukes--Cantor  $(d=3)$,  Kimura-2  $(d=4)$ and  Kimura-3
$(d=5)$. The models are reviewed in Section 2.2.

 Our contribution  is the construction of a whole genome parametric
 alignment in all four models for
 \textit{Drosophila melanogaster} and \textit{Drosophila pseudoobscura}.
 Our methods and computational results  are described in Section~2.
Three biological applications are presented in Section~3.
A discussion follows in Section~4.

\section{From Genomes to Polytopes}

The data we analyzed are the genome sequences of
 \textit{Drosophila melanogaster} (April 2004 BDGP release 4)
 and {\it Drosophila pseudoobscura} (August 2003 freeze 1).
Our main computational result is the construction of a whole genome
parametric alignment for these two genomes. This result depended on
a number of innovations. By adapting existing orthology
mapping methods, we were able to divide the genomes into 1,999,817
pairs of reliably orthologous segments, and among these we
identified 877,982 pairs for which the alignment is uncertain. We
computed the alignment polytopes of dimensions two, three
and four for each of these 877,982 sequence pairs,
and of dimension five for a subset of them.
 The methods are
explained in Section 2.3. The vertices of these polytopes represent
the optimal alignment summaries and the robustness cones. These
concepts are introduced in Section 2.2. Computational results
are presented in Section 2.4.

\subsection{Orthology Mapping}

The {\em orthology mapping problem} for a pair of genomes is to
identify all orthologous segments between the two genomes. These
orthologous segments, if selected so as not to contain genome
rearrangements, can then be globally aligned to each other. This
strategy is frequently used for whole genome alignment
\cite{Gibbs2004, Waterston2002}, and we adapted it for
our parametric alignment computation.

MERCATOR is an orthology mapping program suitable for multiple
genomes that was developed by Dewey et al. \cite{Dewey2005}.
 We applied this program to the {\it D. melanogaster} and {\it D. pseudoobscura}
genomes in order to identify pieces for parametric alignment. The
MERCATOR strategy for identifying orthologous segments is as
follows.  Exon annotations in each genome are translated into amino
acid sequences and then compared to each other using BLAT
\cite{Kent2002}.  The annotations are based on reference gene sets,
such as FlyBase genes \cite{Drysdale2005}, and on {\it ab initio}
predictions.  The resulting exon hits are then used to build a graph
whose vertices correspond to exons, and with an edge between two
exons if there is a good hit. A greedy
algorithm is then used to select edges in the graph that correspond
to runs of exons that are consistent in order and orientation.

The MERCATOR orthology map for  \textit{D.~melanogaster} and {\it
D.~pseudoobscura} has 2,731 segments. However, in order to obtain a
map suitable for parametric alignment, further subdivision of the
segments was necessary. This subdivision
 was accomplished by the additional
step of identifying and fixing exact matches of length at least 10bp.
We did this  using the software MUMMER; see \cite{Delcher1999}.

 We derived 1,116,792 constraints, which are of
four possible types:
\begin{itemize}
\item exact matching non-coding sequences,
\item ungapped high scoring aligned coding sequences,
\item segment pairs between two other constraints where
one of the segments has length zero, so the non-trivial segment must
 be gapped, and
\item single nucleotide mismatches that are squeezed between
other constraints.
\end{itemize}
We then removed all segments where the sequences contained the letter
\texttt{N} (which means the actual sequence is uncertain). 
This process resulted in 877,982 pairs of segments for
parametric alignment. 
The lengths of the {\it D. melanogaster} segments range from 1 to 80,676 base
pairs. The median length is 42bp and the mean length is 99bp.
In all, $90.4\%$ of the {\it Drosophila
  melanogaster} genome and $88.7\%$ of the {\it Drosophila
  pseudoobscura} genome were aligned by our method.

\subsection{Models, Alignment Summaries, and Robustness Cones}

For each of the 877,982 pairs of orthologous segments, we
constructed all optimal Needleman--Wunsch alignments, with
respect to various scoring schemes
which are derived from pair hidden Markov models (PHMMs).
We considered models with 2,3,4, and 5 parameters.
See \cite[\S 4.1]{Durbin1998} for a
review of PHMMs and their relationship to the scoring
schemes typically used for aligning DNA sequences. In what
follows, we only refer to the scores, which are the logarithms of
certain ratios of the parameters of the PHMM.
Our four models are specializations
of the general 33 parameter model \cite[\S 2.2]{ASCB2005}
that incorporates mutations, insertions, and deletions of DNA sequences.
 It is customary to reduce the dimension by
 assuming that many of the 33 parameters are equal to each other.

The {\em basic model}, discussed in
Section~\ref{Sect:Introduction}, has three natural parameters,
namely, $M$ for match, $X$ for mismatch and $S$ for space. If the
numbers $M,$ $X$ and $S$ are fixed, then we seek to maximize $\, M
\cdot m \, + \, X \cdot x \, + S \cdot s$, where $(m,x,s)$ runs over
the summaries of all alignments. In light of the relation
(\ref{linrel}), this model has only two free parameters and there is
no loss of generality in assuming that the match score $M$ is zero.
From now on we set $M=0$ and we take  $X$ and $S$ as the free
parameters. We define the {\em 2d alignment summary} to be the pair
$(x,s)$.

Following the convention of \cite[\S 2.2]{ASCB2005}, we summarize a scoring scheme
with a $5 \times 5$-matrix $w$ whose rows and columns are both
indexed by \texttt{A}, \texttt{C}, \texttt{G}, \texttt{T}, and
\texttt{-}. The matrix $w$ for
the basic model is the leftmost matrix in Table 2, and it
corresponds to the \emph{Jukes--Cantor model} of DNA sequence
evolution.

Our {\em 3d model} is the most commonly used
scoring scheme for computing alignments.
This model includes the number $g$ of gaps. A {\em gap} is a
complete block of spaces in one of the aligned sequences; it either
begins at the start of the sequence or is immediately preceded by a
nucleotide, and either follows the end of the sequence or is
succeeded by a nucleotide. The {\em 3d alignment summary} is the
triple $(x,s,g)$.
 The score for a gap, $G$, is known as the {\em affine gap penalty}.
If $X,$ $S$ and $G$ are fixed, then the alignment problem is to
maximize $\, X \cdot x + S \cdot s + G \cdot g\,$ where $(x,s,g)$
runs over all 3d alignment summaries. The parametric version is
implemented in XPARAL. Introducing the gap score $G$ does not affect
the matrix $w$ which is still the leftmost matrix in Table~2.

Our {\em 4d model} is derived from the {\em Kimura-2 model} of
sequence evolution.
 The {\em 4d alignment summary} is the vector $(x,y,s,g)$
where $s$ and $g$ are as above,
$x$ is the number of {\em transversion mismatches} (between a purine and a pyrimidine or
vice versa) and $y$ is the number of {\em transition mismatches}
(between purines or between pyrimidines). The four parameters are $X$, $Y$, $S$, and $G$. The
matrix $w$ of scores, as specified in \cite[(2.11)]{ASCB2005},
 is now the middle matrix in Table~2.

\begin{table}
\label{threematrices}
$$  \begin{pmatrix}
0 & X & X & X & S \\
X & 0 & X & X & S \\
X & X & 0 & X & S \\
X & X & X & 0 & S \\
S & S & S & S &  \end{pmatrix},\,\,\,
\begin{pmatrix}
0 & X & Y & X & S \\
X & 0 & X & Y & S \\
Y & X & 0 & X & S \\
X & Y & X & 0 & S \\
S & S & S & S &  \end{pmatrix} , \,\,\,
 \begin{pmatrix}
0 & X & Y & Z & S \\
X & 0 & Z & Y & S \\
Y & Z & 0 & X & S \\
Z & Y & X & 0 & S \\
S & S & S & S &  \end{pmatrix}
$$
\vskip -.3cm \caption{The Jukes--Cantor matrix, the Kimura-2 matrix,
and the Kimura-3 matrix.
 These three matrices correspond to
 JC69, K80 and K81 in the {\em Felsenstein
hierarchy} \cite[Figure 4.7]{ASCB2005} of probabilistic models for
DNA sequence evolution.}
\end{table}

Our {\em 5d scoring scheme} is derived from the {\em Kimura-3
model}. Here the matrix $w$ is the rightmost matrix in Table~2. The
{\em 5d alignment summary} is the vector $(x,y,z,s,g)$, where $s$
counts spaces, $g$ counts gaps,  $x$ is the number of mismatches
$\ontop{\mathtt A}{\mathtt C}$, $\ontop{\mathtt C}{\mathtt A}$,
$\ontop{\mathtt G}{\mathtt T}$ or $\ontop{\mathtt T}{\mathtt G}$,
$y$ is the number of mismatches $\ontop{\mathtt A}{\mathtt G}$,
$\ontop{\mathtt G}{\mathtt A}$, $\ontop{\mathtt C}{\mathtt T}$ or
$\ontop{\mathtt T}{\mathtt C}$, and $z$ is the number of mismatches
$\ontop{\mathtt A}{\mathtt T}$, $\ontop{\mathtt T}{\mathtt A}$,
$\ontop{\mathtt C}{\mathtt G}$ or $\ontop{\mathtt G}{\mathtt C}$.
Thus, the 5d alignment summaries of the two {\it Drosophila} intron
alignments at the beginning of Section~\ref{Sect:Introduction} are
$\, (4, 10, 9, 9, 8) \,$ and
$\,   (3, 3, 4, 29, 17) $.
Even the 5d model does not encompass
all scoring schemes that are used in practice.
See Section 3.1 for a discussion of the BLASTZ scoring matrix
\cite{Chiaromonte2002} and its proximity to the Kimura-2 model.

Suppose we are given a specific alignment of two DNA sequences. Then
the {\em robustness cone} of that alignment is the set of all
parameter vectors that have the following property: any other
alignment that has a different alignment summary is given a lower
score. As a mathematical object, the robustness cone is an open
convex polyhedral cone in the space $\mathbb{R}^d$ of free
parameters.

An alignment summary is said to be {\em optimal}, relative to one of
our four models, if its robustness cone is not empty. Equivalently,
an alignment summary is optimal if there exists a choice of
parameters such that the Needleman--Wunsch algorithm produces
\emph{only} that alignment summary. Such a parameter choice will be
robust, in the sense that if we make a small enough change in the
parameters then the optimal  alignment summary will remain
unchanged. Each robustness cone is specified by a finite list of
linear inequalities in the model parameters.

For example, consider the first alignment in the Introduction. 
Its 2d alignment summary is the pair $(x,s) =
(23,9)$, labeled ${\bf D}$ in Table 1. The
robustness cone of this summary is the set of all points $(X,S)$
such that the score $\,23 X + 9 S \,$ is larger than the score of all other
alignments summaries other than $(23,9)$. This cone is specified by
the two linear inequalities $S > X$ and $4S < 3X$.  

If we fix two DNA sequences, then the robustness cones of all the
optimal alignments define a partition of the parameter space,
$\mathbb{R}^d$. That partition is called the {\em alignment fan} of
the two DNA sequences. Figure 1 shows the (biologically relevant
part of the) alignment fan of two {\it Drosophila} introns in the 2d
model.  While this alignment fan has only 13 robustness cones,
the alignment fan of the same introns has 76 cones for the 3d
model, 932 cones  for the 4d model, and 10,009 cones  for the 5d
model. These are the vertex numbers in Table 5.

\subsection{Alignment Polytopes}
\label{SubSect:AlignmentPolytopes}

The {\em convex hull} of a finite set $\mathcal{S}$ of points in
$\mathbb{R}^d$ is the smallest convex set containing these points.
It is  denoted ${\rm conv}(\mathcal{S})$ and called a {\em convex
polytope}. There exists a unique smallest subset $\mathcal{V}
\subseteq \mathcal{S}$ for which $\conv(\mathcal{S}) =
\conv(\mathcal{V})$. The points in $\mathcal{V}$ are called the {\em
vertices} of the convex polytope. The vertices lie in
higher-dimensional \emph{faces} on the boundary of the polytope.
Faces include {\em edges}, which are one-dimensional, and  {\em
facets}, which are $(d-1)$-dimensional. Introductions to these
concepts can be found in the textbooks \cite{Grunbaum2003, Preparata1985}. 
By {\em computing the
convex hull} of a finite set $\mathcal{S} \subset \mathbb{R}^d$ we
mean  identifying the vertices and the facets of  $\,{\rm
conv}(\mathcal{S})$ and, if possible, all faces of all dimensions.

Fix one of the four models discussed in Section 2.2.
The {\em alignment polytope} of two DNA sequences is the convex
polytope ${\rm conv}(\mathcal{S}) \subset \mathbb{R}^d,$ where
$\mathcal{S}$ is the set of alignment summaries of all alignments of
these two sequences. For instance, the 3d alignment polytope of two
DNA sequences is the convex polytope in $\mathbb{R}^3$ that is
formed by taking the convex hull of all alignment summaries
$(x,s,g)$.  Figure 2 shows the 3d alignment polytope
for the two sequences in the Introduction.
Its projection onto the $(x,s)$-plane is 
the polygon depicted in Figure 1.

It is a basic fact of convexity that the maximum of a linear
function over a polytope is attained at a vertex. Thus,
an alignment of two DNA sequences is optimal if
and only if its summary is in the set $\mathcal{V}$ of vertices of
the alignment polytope. The Needleman--Wunsch
algorithm efficiently solves the linear programming problem
over this polytope. For instance, for the 3d
model with fixed parameters, the alignment problem is
the linear programming problem
\begin{equation}
\label{myLP}
 \hbox{Maximize} \,\,\,
  X \cdot x + S \cdot s + g \cdot G
\,\,\,\hbox{subject to} \,\,\,(x,s,g) \in
\mathcal{V}.
\end{equation}
For a numerical example consider the
parameter values $X = -200$, $S=-80$
and $G  = -400$, which
represent an approximation of the BLASTZ scoring scheme (Section 3.1).
The solution to (\ref{myLP}) is
attained at the vertex $(x,s,g)=(30,5,2)$
which is the 3d summary of the following alignment
of our two {\it Drosophila} introns 
\begin{small}
\begin{verbatim}
mel GTAAGTTTGTTTATATTTTTTTTTTTTTGAAGTGACAAATAGC--ACTTATAAATATACTTAG
pse GTTCGTTAACACATGAAATTCCATCGCCTGATTGTTCACTATCTAACTAACGAAT---TTTAG
    **  ***     **    **   *      * **   * ** *  *** *  ***    ****
\end{verbatim}
\end{small}
The 3d summary of this alignment is the marked vertex in 
Figure 2.

\begin{figure}[ht]
\label{fig:3dpolytope}
\begin{center}
\includegraphics[scale=1.2]{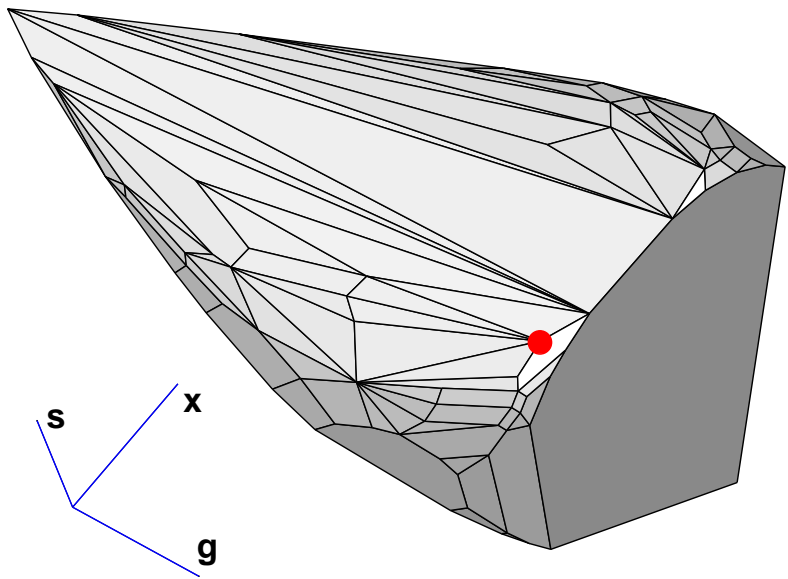}
\end{center}
\caption{The 3d alignment polytope of our two {\it Drosophila} introns
has 76 vertices. The marked vertex 
$(x,s,g)=(30,5,2)$ represents the BLASTZ alignment.
}
\end{figure}

The problem of computing a parametric alignment is now specified
precisely. The input consists of two DNA sequences. The output is
the set of vertices and the set of facets of the alignment polytope
${\rm conv}(\mathcal{S})$. See also \cite[Remark 2.29]{ASCB2005}.
We note that
the robustness cone of an optimal alignment summary is
the {\em normal cone} of the polytope at that vertex. The alignment
fan is the {\em normal fan} of the alignment polytope. See
\cite[p.~61]{ASCB2005} for definitions of these concepts.

We considered two different methods for constructing
all alignment polytopes for the two {\it Drosophila} genomes:
{\em polytope propagation} and {\em incremental convex
hull}.  In our study we found that polytope propagation was
outperformed by the incremental convex hull algorithm, especially
for the higher dimensional models.

We now briefly outline the two methods.  Polytope propagation for
sequence alignment is the Needleman--Wunsch algorithm with the
standard operations of plus and max replaced by Minkowski sum and
polytope merge (convex hull of union).  The polytope propagation
algorithm was introduced in \cite{Pachter2004b, ASCB2005}.

The incremental convex hull algorithm, on the other hand, gradually
builds the alignment polytope by successively finding new optimal
alignment summaries, the vertices of the polytope. In order to find
the new optimal summaries, the algorithm repeatedly calls a {\em
Needleman--Wunsch (NW) subroutine} that is an efficient
implementation of the classical Needleman--Wunsch algorithm. For
fixed values of the parameters, this subroutine returns an optimal
alignment summary. For instance, for the 3d model, the input to the
NW subroutine is a parameter vector $(X,S,G)$ and the output is an
optimal summary $(x,s,g)$.

Suppose we have already found a few optimal alignment summaries, by
running the NW subroutine with various parameter values. We let $P$
be the convex hull of the summaries in $\mathbb{R}^d$, and we assume
that $P$ is already $d$-dimensional.  We maintain a list of all
vertices and facets of $P$. Each facet is either \emph{tentative} or
\emph{confirmed}, where being confirmed means that its affine
span is already known to be a facet-defining hyperplane  of the final alignment polytope.
In each iteration, we pick a tentative facet of $P$ and an outer
normal vector ${\bf U}$ of that facet.  We then call the NW
subroutine with ${\bf U}$ as the input parameter. The output of the
NW subroutine is an optimal summary ${\bf v}$. If the optimal score
${\bf U} \cdot {\bf v}$ equals the maximum of the linear function
${\bf U} \cdot {\bf w}$ over all ${\bf w}$ in $P$ then we declare the facet to
be confirmed. Otherwise, the score ${\bf U} \cdot {\bf v}$ is
greater than the maximum and we replace $P$ by the convex hull of
$P$ and ${\bf v}$. This convex hull computation
utilizes the {\em beneath-beyond construction}
\cite[\S3.4.2]{Preparata1985} which erases some of
the tentative facets of the old polytope and replaces them by new
tentative facets.

The algorithm terminates when all facets are confirmed. The current
polytope $P$ at that iteration is the final alignment polytope. The
number of iterations of this incremental convex hull algorithm
equals the number of vertices plus the number of facets of the final
polytope $P$.  So for a given model, the running time
of the incremental convex hull algorithm scales
linearly in the size of the output.
This was confirmed in practice
 by our computations (Table 3).

Our software, together with 
more details about our incremental convex hull implementation are
available for download at the supplementary website.

Given an alignment polytope, there are various subsequent
computations one may wish to perform. For instance,
we may be interested in  the robustness
cones at the vertices. In order to get an irredundant
inequality representation of a robustness cone, it suffices to know
the edges emanating from the corresponding vertex.  Thus it is
useful to also compute the edge graph of each of our 
polytopes.

\begin{table}
\label{Table:RunningTimes}
\begin{center}
\begin{tabular}{c | c}
& running time (in seconds)\\
\hline
d=2 & $4.52\cdot 10^{-2} + 6.16 \cdot 10^{-7} Vll'$\\
d=3 & $4.76 \cdot 10^{-2}  +  9.28 \cdot 10^{-7} Vll'  +  2.41 \cdot 10^{-7} Fll'$\\
d=4 & $1.05 \cdot 10^{-1}  +  9.53 \cdot 10^{-7} Vll'  +  3.84 \cdot
10^{-7}  Fll'$\\
d=5 &  $16.0  +  1.20 \cdot 10^{-6}  Vll'  +  5.66 \cdot 10^{-7}  Fll'$
\end{tabular}
\end{center}
\caption{Observed running times of the incremental convex hull algorithm
for computing alignment polytopes. Here, $V$
is the number of vertices of a polytope, $F$ the number of facets,
and $l,l'$ are the sequence lengths. The given functions are a best-fit
estimation from a sample of 764 out of the 877,982 sequence pairs. In
particular, 90\% of the actual measured running times from this
sample were within 10\% of this estimation. The running times are on a
2.5 GHz machine.}
\end{table}

\subsection{Computational Results}

Using our implementation of the incremental convex hull algorithm
described above, we computed the 2d, 3d, and 4d polytopes for each
of the 877,982 segment pairs. We also computed 5d polytopes in many
cases.
These polytopes are available for
downloading and viewing at the supplementary website
$$ \hbox{\url{http://bio.math.berkeley.edu/parametric/}.}$$

We empirically determined the
expected CPU time to construct alignment polytopes.
The results are reported in Table 3.
As expected, the running time of the incremental convex hull
algorithm scales linearly with the number of vertices plus facets.
The running time of a single Needleman--Wunsch subroutine
call scales linearly with the product $ll'$
of the sequence lengths $l$ and $l'$.

In order to effectively compute these polytopes, not only must we
have an algorithm which runs quickly as a function of the number of
vertices and facets, but the number of vertices and facets must
themselves be small. The theoretical bounds discussed in
the Introduction ensure that these numbers grow
polynomially, for any fixed $d$. In our computations
we found that the numbers of vertices and facets 
of alignment polytopes are quite manageable
even in dimensions $4$ and $5$. Averages of the
numbers we actually observed are reported in Table 4.

\begin{table}
\begin{center}
\begin{tabular}{r | c c c c}
$d$ & 2 & 3 & 4 & 5\\
\hline
Average $V$ & 5.8 & 47.8 & 580.8 & 6406.0\\
Average $F$ & 5.8 & 47.1 & 859.4 & 18996.5\\
\end{tabular}
\end{center}
\caption{Averages of the  number $V$ of vertices
and the number $F$ of facets of alignment polytopes.
 These averages are from the same
sample as in Table 3.  The average of the sum $l+l'$ of the
sequence lengths for this sample was 82.7.}
 \end{table}

We conclude our summary of computational results with a look
at the alignment polytopes of the orthologous pair of introns
in the Introduction.
The sequence lengths are $l = 60 $
and $l' = 61$. The 2d alignment polytope
is shown in Figure 1. The 3d alignment polytope is shown in Figure 2.  Table 5
reports statistics
for the 2d, 3d, 4d, and 5d alignment
polytopes for this sequence pair.

\begin{table}
\begin{center}
\begin{tabular}{r | c c c c}
$d$ & 2 & 3 & 4 & 5\\
\hline
\# of vertices & 13  & 76  & 932  & 10009\\
\# of edges    & 13  & 159 & 3546 & 66211\\
\# of 2d faces & --- & 85  & 4208 & 139723\\
\# of 3d faces & --- & --- & 1594 & 118797\\
\# of 4d faces & --- & --- & ---  & 35278 \\
Avg. \# of edges per vertex & 2 & 4.2 & 7.6 & 13.2
\end{tabular}
\end{center}
\caption{Face numbers of the alignment polytopes for the intron
sequences from the beginning of the Introduction. The average number
of edges containing a vertex is the average number of linear
inequalities bounding a robustness cone.}
\end{table}

\section{From Polytopes to Biology}

We describe three applications of our whole genome parametric
alignment. First, we discuss how alignment polytopes are useful for
parameter selection, and we assess the BLASTZ alignment of \textit{D. melanogaster} and
\textit{D. pseudoobscura}. We then revisit the cis-regulatory
element study in \cite{Richards2005}, and we determine alignments that 
identify previously missed conserved binding sites.
Finally, we examine the problem of branch length estimation and provide a
quantitative analysis of the dependence of branch length estimates on alignment parameters.

\subsection{Assessment of the BLASTZ alignment}

A key problem in sequence alignment is to determine appropriate
parameters for a scoring scheme. The standard approach is 
to select a model and 
then identify parameter values that are effective in producing 
alignments that correctly align certain features. For example, 
the BLASTZ scoring matrix \cite{Chiaromonte2002} was optimized 
for human-mouse alignment by finding parameters that were effective
in aligning genes in the HOXD region. The 
BLASTZ scoring scheme is given by a scoring matrix called HOXD70 \cite{Chiaromonte2002},
$$  \bordermatrix{ & A & C & G & T \cr
A & 91 & -114 & -31 & -123 \cr
C & -114 & 100 & -125 & -31 \cr
G & -31 & -125 & 100 & -114 \cr
T & -123 & -31 & -114 & 91 \cr
}, 
$$
together with a space score of -30, and a gap score of -400. 
Although the HOXD70 matrix has six distinct entries,
it can be approximated by a Kimura-2 matrix (Table 2),
since 91 is close to 100, and
114 is close to 123 and 125.

The alignment polytope of a pair of DNA sequences is a representation
of all possible alignments organized according to
a scoring scheme. Thus, our results and methodology make it possible, for the first time,
to identify parameters that
are guaranteed to optimize the alignment according to desired criteria.
Moreover, our results offer biologists a mathematical tool for
systematically assessing whether a proposed single alignment is suitable
for its intended purpose.

We initiated such a study for
 the BLASTZ \cite{Schwartz2003} alignment of \textit{D.
melanogaster} and \textit{D. pseudoobscura}, which is available at {\tt
http://genome.ucsc.edu/}. This alignment is widely used
by biologists who study {\it Drosophila}.
Although the BLASTZ alignment procedure
is based on an initial ``seeding'' procedure (similar to 
our identification of constrained segment pairs), the alignments
are then constructed using the Needleman--Wunsch algorithm with 
the HOXD70 matrix. 

Recall that our orthology map consisted of 1,999,817 segment pairs: 
1,116,792 consisted of segments for which we fixed the
alignment (constrained segment pairs) and 883,025 were
unconstrained segment pairs for which 
we constructed alignment polytopes. 
We found that the BLASTZ alignment agreed with 623,710 of our 
unconstrained segment pairs, of which 622,173 did not contain {\tt N}'s. 
For each of these  622,173 
segment pairs, we computed the 2d, 3d and 4d alignment summaries
for the BLASTZ alignments, and we determined whether or not
they are optimal for some choice of model parameters.

We found that 269,186 ($43.3\%$) of the BLASTZ alignments 
are vertices of the 3d polytope, but not the 2d
polytope, and 201,982 ($32.5\%$) are vertices of both the 2d and 3d polytopes.
Only 151,004 ($24.3\%$)
of the  BLASTZ alignments are not vertices of either the 2d or 3d
alignment polytopes. 
In summary, our computations show that
$32.5\%$ of the BLASTZ alignments correspond to vertices of
the 2d polytope and $75.7\%$ correspond to vertices of the 3d
polytope. These numbers are even higher for the 4d and 5d 
alignment polytopes.

Curiously, there is precisely one sequence pair
where the BLASTZ alignment is a vertex of the 2d 
polytope but not the 3d polytope. This alignment is
 \begin{verbatim}
mel AGCCGAACCGGATATCCAGGCCGAGGCC
pse GCCAGAGCCGGA-GCCTGAGCCGGAG--
      * ** *****   *    ***  *
 \end{verbatim}
\vskip -.5cm
The 3d summary of this alignment is $(11,3,2)$, which
is the midpoint of the edge
with vertices $(11,3,1)$ and $(11,3,3)$ on
the 3d polytope.
 Hence this alignment
is not optimal for any choice of
parameters $(X,S,G)$. However, it is
optimal for the 2d model since the edge maps onto the
vertex $(11,3)$ of the 2d polygon.

Our results show that not only does the BLASTZ alignment agree 
well with our constrained segment pairs, but, even on the unconstrained
segment pairs, the BLASTZ alignments are mostly vertices of 
the three dimensional polytopes. This suggests that there may 
be a statistical advantage to working with 
one of the lower dimensional models, and 
also indicates that the polytopes may be useful for finding 
parameters. We illustrate this point of view in the next section,
where we identify vertices in the alignment polytope (and therefore
parameter robustness cones) that are suitable for the alignment of
cis-regulatory elements. Any user of the BLASTZ alignments
may now use the alignment polytopes we provide in order to assess whether or not
the fixed choice of the HOXD70 matrix is the right one for their
particular biological application.

\subsection{Conservation of cis-regulatory elements}

A central question in comparative genomics is the extent of
conservation of cis-regulatory elements, and the implications 
for genome function and evolution. Using our parametric alignment, 
we discovered that cis-regulatory elements may be more conserved
between {\it D. melanogaster} and {\it D. pseudoobscura} than
previously thought. Specifically, we used our alignment polytopes to examine
the 
degree of conservation for 1346 transcription factor binding sites
\cite{Bergman2005} available at {\tt www.flyreg.org}
(we excluded 16 sites which were located in segment pairs containing $N$s). 
The 1346 sites
include the 142 sites examined by Richards et al.
\cite{Richards2005} in their comparison of \textit{D. pseudoobscura} and
\textit{D. melanogaster}. 

Specifically, for each of the 1346 elements, we identified the
orthologous segment pairs from our orthology map that contained the
elements. We then extracted the polytopes from our whole genome parametric alignment.
For each polytope, we determined an optimal
alignment for which the number of matching bases of the
corresponding element was maximized.

As an example consider the transcription
factor {\tt Adf1}. It binds to a cis-regulatory element at
chr3R:2,825,118-2,825,144 in {\it D. melanogaster}
(\texttt{Adf1->} \texttt{Antp:06447} in the flyreg database). The BLASTZ
alignment for this element~is
\begin{verbatim}
mel TGTGCGTCAGCGTCGGCCGCAACAGCG 
pse TGT-----------------GACTGCG
    ***                  ** ***
\end{verbatim}
This alignment suggests that the {\it D. melanogaster} cis-regulatory 
element is not conserved in {\it D. pseudoobscura}. However, there are
many optimal alignments which indicate that this element is conserved.
Examining our constrained segment pairs, we found that the prefix {\tt TGTG} was at the end of a 13bp exact
match. The remaining D. melanogaster 
element was part of a segment pair which has 
813 distinct optimal alignments in the 3d model.
Among these, we found the following alignment with parameters $G=-3, S=-8, X=-18$:
\begin{verbatim}
mel TGTG----CGTCAGC--G----TCGGCC---GC-AACAG-CG
pse TGTGACTGCG-CTGCCTGGTCCTCGGCCACAGCCAAC-GTCG
    ****    ** * **  *    ******   ** *** * **
\end{verbatim}
Note that we include the {\tt TGTG} prefix in order to show a complete 
alignment of the cis-regulatory element. The second alignment
has 24 matches instead of the BLASTZ alignment with 8. 
The number of matches can be used to calculate the {\em percent identity} 
for an element as follows:
\[ \hbox{percent identity} \quad = 
\quad 100 \times \frac{\# \hbox{matches}}{\# \hbox{bases in element}}. \]
Percent identity was used in \cite{Richards2005} as a criterion
for determine whether binding sites are conserved. 
The BLASTZ alignment has $30\%$ identity and the alignment
with 24 matches has $89\%$ identity. It is 
an optimal alignment with the highest possible percent identity.
Examining all 813 optimal alignments, it appeared to us that 
the following alignment (obtained with 
$G=-882, S=-87, X=-226$)
is more reasonable, even though it has
a lower percent identity ($67\%$):
\begin{verbatim}
mel TGTGCGTCAGC------GTCGGCCGCAACAGCG
pse TGTGACTGCGCTGCCTGGTCCTCGGCCACAGC-
    ****  *  **      ***  * ** *****
\end{verbatim}
This alternative alignment suggests that the percent identity criterion 
may not be the best way to judge the conservation of elements.
 Regardless, 
we believe our parametric alignment indicates that in this particular case, 
the {\it D. melanogaster} cis-regulatory element is likely to have been
conserved in {\it D. pseudoobscura}. 

 \begin{table}[!h]
 \label{CREtable}
 $$ \begin{matrix}
  && 2d && 3d\\
 \hbox{Mean \% id opt param} && 80.4 && 85.1\\
 \hbox{Mean \% id fix param} && 79.1 && -
 \end{matrix}
 $$
 \caption{Cis-regulatory element conservation.}
 \end{table}

Our overall results are summarized in Table 6. We found that parameters
can be chosen so as to significantly increase the number of matches
for cis-regulatory elements. The ``opt param'' row in the table
shows results for the case where parameters were chosen separately for 
each segment pair so as to maximize the \% identity of the 
cis-regulatory elements. The ``fix param'' row shows results
when one parameter was selected (optimally) for all 
segment pairs simultaneously (this was only computed for 
the 2d model). Note that
the mean per site \% identity reported in \cite{Richards2005} was
$51.3\%$, considerably lower than what we found 
using the whole genome parametric alignment
(even for the 2d model).

Our results seem to indicate that
cis-regulatory elements are more conserved between {\it D. melanogaster} and {\it D. pseudoobscura} 
than previously thought. The alignment polytopes should be a useful 
tool for further investigation of the extent of conservation of 
cis-regulatory elements among the {\it Drosophila} genomes.

\subsection{The Jukes--Cantor distance function}

An important problem in molecular evolution is the estimation of 
branch lengths from aligned genome sequences.  A widely used
method for estimating branch lengths is based on
the Jukes--Cantor model of evolution \cite{Jukes1969}.
Given an alignment of two sequences of lengths $l,l'$, with 2d
alignment summary $(x,s)$, one computes the
{\em Jukes--Cantor distance} of the two genomes as follows:
\[ d_{JC}(x,s) \quad = \quad -\frac{3}{4}{\rm log} \left( 1 - \frac{4}{3}\left(\frac{2x}{l+l'-s}\right) \right). \]
See \cite[Proposition 4.6]{ASCB2005} for a derivation of this expression
which is also known as the {\em Jukes--Cantor correction} of the two
aligned sequences. The Jukes--Cantor distance can be interpreted as 
the expected number of mutations per site.

Since the Jukes--Cantor distance $d_{JC}(x,s)$ depends on the underlying pairwise
sequence alignment summary, which in turn depends on the alignment parameters,
it is natural to ask how the branch length estimate depends on the
parameters in a 2d scoring scheme. We therefore introduce the
{\em Jukes--Cantor distance function} which is the function $\,JC:\mathbb{R}^2 \rightarrow [0,\infty)\,$
given by $(X,S) \mapsto d_{JC}(\hat{x},\hat{s})$
where $(\hat{x},\hat{s})$ is the alignment summary maximizing $X \cdot x + S \cdot s$. 

We computed the Jukes--Cantor distance function
$JC$ for the entire genomes of 
{\it D. melanogaster} and {\it D. pseudoobscura}. As the result of this computation,
we now know the Jukes--Cantor distances  for all whole genome alignments 
which are optimal for some choice of biologically reasonable parameters $(X,S)$.

The notion of ``optimal''  used here rests on the following precise definitions.
Given parameters $(X,S)$,  the optimal 2d alignment summary $(x,s)$ for the
two genomes
is the sum of the optimal summaries of all 877,982
unconstrained segment pairs plus the
sum of the alignment summaries of
the non-coding constrained segment pairs (which do not depend on the parameters).
We determined that the constrained segment pairs contained 91,355
mismatches and 16,339,305 matches.
The {\em genome alignment polytope} is the Minkowski sum
of the 877,982 alignment polytopes. The vertices of the genome
alignment polytope correspond to optimal summaries of whole genome alignments.

We computed the genome alignment polytope for the 2d model.
Remarkably, this convex polygon, which is the Minkowski sum of
close to one million small polygons as in Figure 1, 
was found to have only 1,183 vertices. Moreover,
of the 1,183 vertices of the genome alignment polytope,
only 838 correspond to biologically reasonable parameters $(X<0, \, 2S<X)$.
The finding that there are so few vertices constitutes a striking experimental validation
of Elizalde's Few Inference Functions Theorem \cite[\S 9]{ASCB2005}
in the context of real biological data.

\begin{figure}
\begin{center}
\includegraphics[scale=0.66]{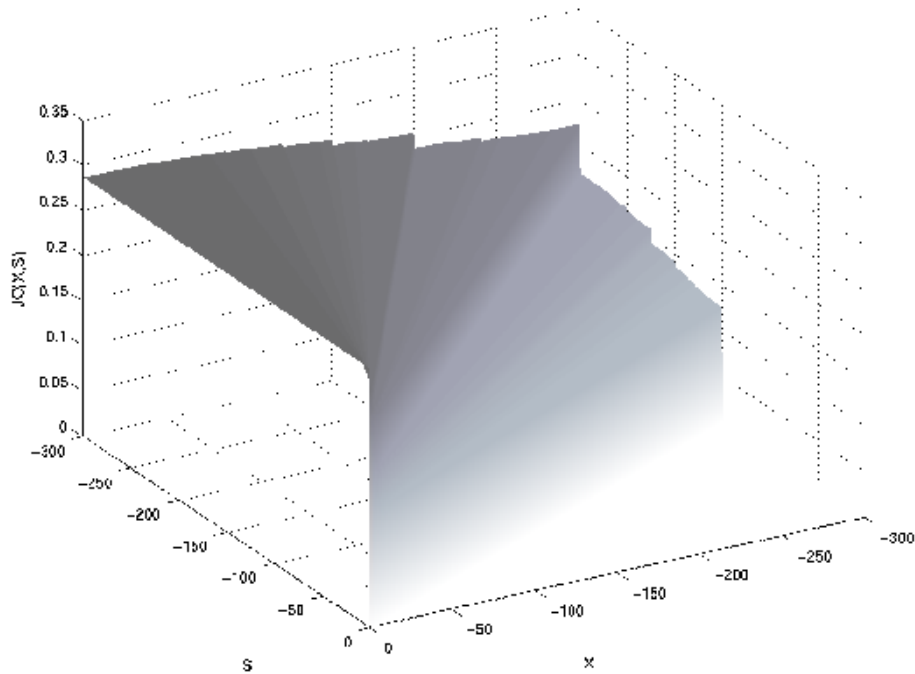}
\end{center}
\caption{The Jukes--Cantor distance function of two
Drosophila genomes.}
\end{figure}

The Jukes--Cantor distance function $JC$ of
{\it D. melanogaster} and {\it D. pseudoobscura}
is a piecewise constant function on the $(X,S)$-plane.
Indeed, $JC$ is constant on the cones in the normal fan
of the genome alignment polygon. Note that $JC$ 
is undefined when $(X,S)$ is perpendicular to one
of the 1,183 edges of the genome alignment polygon.
On such rays, the Jukes--Cantor distance function jumps between
its values on the two adjacent cones in the normal fan.

The graph of the Jukes--Cantor distance function  is shown in
Figure 3. The function ranges in value from $0.1253$ to $0.2853$, is monotonically decreasing as a function of $S$, and monotonically increasing as a function of $X$.
We found it interesting that at the line $X=S$, there is a large ``Jukes--Cantor jump'' where 
the value of the function increases from $0.1683$ to $0.2225$.  

The Jukes--Cantor distance function is a new tool for parametric  
reconstruction of phylogenetic trees.
Instead of estimating a single distance between each pair of genomes in a multiple species 
phylogenetic reconstruction, one can now evaluate the Jukes--Cantor function at
vertices of the Minkowski sum of the whole genome alignment polytopes. These can 
be used for parametric phylogenetic reconstruction using 
distance-based methods such as neighbor joining.

\section{Discussion}

The summary of a pair of aligned sequences is a list of numbers
that determine the score for a scoring
scheme. The alignment polytope is a geometric representation of the summaries
of all alignments. It is an organizing tool for
working with all alignments through their summaries.
We view the Needleman--Wunsch algorithm as a fast
subroutine for finding vertices of the alignment
polytope. The construction of alignment polytopes is useful
for biological studies based on sequence alignments where the
conclusions depend on parameter choices.

We have highlighted three biological applications of our parametric alignments, namely
the problem of parameter selection for sequence alignment, functional element conservation, 
and estimation of evolutionary rate parameters. In each case, our perspective 
suggests new directions for further research. 

Alignment polytopes offer a systematic approach to solving the parameter selection problem.
Although this paper did not address statistical 
aspects of parameter selection, we wish to emphasize
that the vertices of the polytopes represent {\em maximum a posteriori}
estimates of alignments for pair hidden Markov models. 
Our polytopes provide
a setting for developing statistically sound methods for parameter selection that are not
dependent on pre-existing alignments. 

Our results on cis-regulatory elements show that they may be significantly more conserved than 
previously thought, and suggest that, in contrast to the analysis of ultra-conserved elements, 
sequence alignment procedures can be crucial in the analysis of certain functional elements. 
The ongoing {\it Drosophila} genome projects (consisting of sequencing 12 genomes of related species)
offer an extraordinary opportunity for extending our study and further 
exploring cis-regulatory element conservation. This leads to the question of multiple alignment, 
which we have not addressed in this paper, but which we believe presents a 
formidable and important challenge in biological sequence analysis. In particular, 
it will be interesting to explore the geometric point of view we have proposed
and to develop parametric algorithms for multiple sequence alignment. 

The Jukes--Cantor distance function, computed here for the first time,
will be important for determining the robustness of evolutionary studies 
based on sequence alignments.
Estimates of the neutral rate of evolution, which are crucial for comparative 
genomics studies, can hopefully 
be improved and further developed using our mathematical tools. 
The Jukes--Cantor distance function opens up the possibility of parametric distance-based
phylogenetic reconstruction. An immediate next step is the extension of our results
to other phylogenetic models.

The construction of millions of alignment polytopes from two
{\it Drosophila} genomes has revealed mathematical insights that 
should be explored further. For example, we observed empirically that
alignment polytopes have few facets. Although we have not explored the
combinatorial structure of alignment polytopes in this paper, 
this offers a promising direction for 
improving our parametric alignment algorithms and is an interesting direction for 
future research.

\section{Acknowledgments}
Colin Dewey was supported by the NIH (HG003150), Peter Huggins was
supported by an ARCS Foundation fellowship, and Kevin Woods was
supported by the NSF (DMS-040214). 
Bernd Sturmfels was supported by the NSF (DMS-0456960) and 
Lior Pachter was supported by the NIH (R01-HG2362-3 and HG003150) and an NSF CAREER award  (CCF-0347992).

\bibliographystyle{unsrt}
\bibliography{ploscb}
\end{document}